\documentclass[twocolumn,showpacs,showkeys,preprintnumbers,amsmath,amssymb]{revtex4}

\usepackage{graphicx}
\usepackage{dcolumn}
\usepackage{bm}
\usepackage{epsfig}

\begin{document}
\preprint{New Journal of Phys. - IST/IPFN 2009-Pinheiro}

\title[]{On Newton's Third Law}

\author{Mario J. Pinheiro} \email{mpinheiro@ist.utl.pt}
\affiliation{Department of Physics and Institute for Plasmas and
Nuclear Fusion, Instituto Superior T\'{e}cnico, Av. Rovisco Pais,
1049-001 Lisboa, Portugal}

\thanks{We would like to thank partial support from the Funda\c{c}\~{a}o
para a Ci\^{e}ncia e a Tecnologia (FCT) and the Rectorate of the
Technical University of Lisbon.}


\pacs{04.20.Cv; 03.50.De; 05.20.-y; 05.70.Ln}

\keywords{Fundamental problems and general formalism; Classical electromagnetism, Maxwell equations;Classical statistical mechanics; Nonequilibrium and irreversible thermodynamics}

\date{\today}%
\begin{abstract}
The law of action-reaction is thoroughly used in textbooks to derive the conservation laws of
linear and angular momentum, and it was considered by Ernst Mach the the cornerstone of physics.
We give here a background survey of several questions raised by the action-reaction law, and in particular, the role of the physical vacuum is shown to provide an appropriate framework to clarify the occurrence of possible violations of the action-reaction law. It is also obtained an expression for the general linear momentum of a body-particle in the context of statistical mechanics. It is shown
that Newton's third law is not verified in systems out of equilibrium due to an additional entropic gradient term present in
the particle's momentum.
\end{abstract}

\maketitle

\section{Introduction}

The law of action-reaction, or Newton's third law~\cite{Newton}, is
thoroughly used in textbooks to derive the conservation laws of
linear and angular momentum. Ernst Mach considered the third law as
``his most important achievement with respect to the
principles"~\cite{Mach_01,Jammer}. However, the reasoning used
primarily by Newton applies to point particles without structure and
is not concerned with the motion of material bodies composed with a
large number of particles, in or out of thermal equilibrium.

Ernst Mach sustained that the concept of mass and Newton's third law
were redundant; that in fact it should be enough to define
operationally the mass of a given body as the unit of mass to be
sure that ``If two masses 1 and 2 act on each other, our very
definition of mass asserts that they impart to each other contrary
accelerations which are to each other respectively as
2$:$1"~\cite{Mach_01}. Yet philosophy has delivered us extraordinary
new insights to a basic understanding of the underlying physics of
force. For example, F\'{e}lix Ravaisson~\cite{Ravaisson} in the XIX
century sustained that within the realm of the inorganic world
action-equals-reaction; they are the same act perceived by two
different viewpoints. But in the organic world, whenever more
complex systems are at working, ``Ce n'est pas assez d'un moyen
terme indiff\'{e}rent comme le centre des forces oppos\'{e}es du
levier; de plus en plus, il faut un centre qui, par sa propre vertu,
mesure et dispense la force"~\cite{Note}. So, there is in Nature the need of an ``agent" that control
and deliver the action from one body to another and this is, as we
will see, the role of the physical vacuum, or barely just the
environment of a body.

We can find in Cornille~\cite{Cornille} a review of applications of
action-reaction law in several branches of physics. In addition,
Cornille introduced the concepts of spontaneous force (obeying to
Newton's third law) and stimulated force (which violates it).

In this paper we intend to show that generally in any system out of
equilibrium, when entropy is velocity-dependent, Newton's third law
is violated. The need for re-examination of this problems is
pressing since long-term exploitation of the cosmos face serious
difficulties due to the outdated spacecraft technologies mankind
possess.

Sec. II discusses the general issues in mechanics and electromagnetism related to the action-to-reaction law. Sec. III discusses the possible role of physical vacuum as a third agent that might explain action-to-reactions law violations. Secs. IV and V discusses the intrinsic violation of Newton's third law in systems out-of-equilibrium. Sec. VI presents the conclusions that follow logically from the previous discussion.

\section{Background survey}

The usual derivation of the laws governing the linear and angular
momentum presented in textbooks is as follows. The equation of
motion of the $i$th particle is given by:
\begin{equation}\label{eq0}
\mathbf{F}_i + \sum_{j \neq i} \mathbf{F}_{ij} = \frac{d
\mathbf{p}}{d t},
\end{equation}
where $\mathbf{F}_i$ is an external force acting on the $i$ particle
and $\mathbf{F}_{ij}$ represents the internal force exerted on the
particle $i$ by the particle $j$. In the case of central forces the
relation $\mathbf{F}_{ij}=-\mathbf{F}_{ji}$ is verified, a
manifestation of Newton's third law. Summing up over all the
particles belonging to the system we have from Eq.~\ref{eq0}:
\begin{equation}\label{eq01}
\sum_i \mathbf{F}_i=\sum_i \frac{d \mathbf{p}_i}{d t}.
\end{equation}
Podolsky~\cite{Podolsky} called
attention to the discrepancies obtained using directly Newton's
second law, or using instead the invariance of the lagrangian under
rotations. In the case of non-central forces, like a system
subject to a potential function of the form $V=r^{-1} \cos
\vartheta$, we might expect a deviation from Newton's third law.
Indeed, angle-dependent potentials, long-range (van der Waals)
forces describe rigorously the physical properties of molecular
gases. We can wonder from which mechanism it comes the unbalance of forces.

We might expect that thermodynamics and statistical mechanics provide a better
description of macroscopic matter. The internal energy and in
particular the average total energy of a system $\overline{E}=\sum_i
U_i$, which includes summing up all the particles constituting the
system and all storage modes, plays a fundamental role together with
an equally fundamental, although less understood function, the
system entropy. Interesting enough a microscopic model of friction have shown that
the irreversible entropy production is drawn from the increase of
Shannon information~\cite{Diosi_02}.

This question is related to the fundamental one still not answered
in physics and biophysics: how chaos in various natural systems can
spontaneously transform to order? The observation of various
physical and biological systems shows that a feedback is onset
according to: ``The medium controls the object-the object shapes the
medium"~\cite{Tsyganov_91}. At the microscopic level a large class
of systems generate directed motion through the interaction of a
moving object with an inhomogeneous substrate periodically
structured~\cite{Popov_02}. This is the ratchet-and-pawl principle.

The apparent violation of Newton's third law that we can find in
some systems, e.g., when two equal charged bodies having equal
velocities in magnitude and opposing directions, is well-known. The
Lorentz's force applied to both charges do not cancel each other
since the magnetic forces do not act along a common line (see also
the Onoochin's paradox~\cite{McDonald}). The paradox is solved
introducing the electromagnetic momentum $[\mathbf{E} \times
\mathbf{H}]/ c^2$ (values in SI units will be used throughout the text)~\cite{Keller_42}.

In the domain of astrophysics, the same problem appears again. For instance,
based on unexplained astrophysical observations, such as the high
rotation of matter around the centers of galaxies, it was proposed a
modification of Newton's equations of dynamics~\cite{Milgrom_83},
while more recently a new effect was reported, about the possibility
of a violation of Newton's second law with static bodies experimenting
spontaneous acceleration~\cite{Ignatiev_07}. In the frame of statistical mechanics, studying the effective
forces between two fixed big colloidal particles immersed in a bath
of small particles, it has been shown that the nonequilibrium force field is
nonconservative and violates action-reaction law~\cite{Likos}.

An ongoing debate on the validity of electrodynamic force law is
still raging, with experimental evidence that Biot-Savart law does
not obey action-reaction law (see Ref.~\cite{Graneau_82,Graneau_01}
and references therein). The essence of the problem stands on two
different laws that exist in magnetostatics giving the force between
two infinitely thin line-current elements $d \mathbf{s}_1$ and
$d\mathbf{s}_2$ through which pass currents $i_1$ and $i_2$. The
{\it Amp\`{e}re's law} states that this force is given by:
\begin{equation}\label{eq2}
d^2 \mathbf{F}_{2,A}=-\frac{\mu_0 i_1 i_2}{4
\pi}\frac{\mathbf{r}_{12}}{r^3_{12}}[ 2(d\mathbf{s}_1 \cdot
d\mathbf{s}_2) - \frac{3}{r^2_{12}}(d\mathbf{s}_1 \cdot
\mathbf{r_{12}})(d\mathbf{s}_2 \cdot \mathbf{r}_{12}) ].
\end{equation}
This means that the force between two current elements depended not
only on their distance, as in the inverse square law, but also on
their angular position (in particular, implicating the existence of
a longitudinal force, confirmed experimentally by Saumont~\cite{Saumont_68} and Graneau~\cite{Graneau_87}, and discussed by Costa de Beauregard~\cite{Beauregard_93}). The other force generally considered is given
by the {\it Biot-Savart law}, also known as the Grassmann's equation
in its integral form:
\begin{equation}\label{eq3}
d^2 \mathbf{F}_{2,BS} = - \frac{\mu_0 i_1 i_2}{4 \pi}
\frac{1}{r^3_{12}} [ (d\mathbf{s}_2 \times (d\mathbf{s}_1) \times
\mathbf{r}_{12})].
\end{equation}
Here, $\mathbf{r}_{12}$ is the position vector of element 2 relative
to 1. While Amp\`{e}re's law obeys Newton's third law, Biot-Savart
law does not obey it (e.g.,
Ref.~\cite{Christodoulides_88,Graneau_94,Valverde1,Valverde2}). The theory developed by
Lorentz was criticized by H. Poincar\'{e}~\cite{Poincare_00}, because it sacrificed action-to-reaction law.

The problem of linear momentum of stationary system of charges and
currents is faraway of a consensus too. Costa de
Beauregard~\cite{Costa_67} pointed out a violation of the
action-reaction law in the interaction between a current loop I
flowing on the boundary of area $\mathbf{A}$ with moment
$\mathbf{\mathfrak{M}}=I \mathbf{A}$ and an electric charge
concluding that when the moment of the loop changes in the presence
of an electric field a force must act on the current loop, given by
$\mathbf{F}=[\mathbf{E} \times \mathbf{\mathfrak{\dot{M}}}]/c^2$.
Shockley and James~\cite{Shockley_67} attribute $\mathbf{F}$ to a
change in ``hidden momentum" $\mathbf{G}_l=-[\mathbf{E} \times
\mathbf{\mathfrak{M}}]/c^2$ carried within the current loop by the
steady state power flow necessary to balance the divergence of
Poynting's vector. The total momentum is $\mathbf{p}=\mathbf{G}_l +
\mathbf{G}_b$, where $\mathbf{G}_b=m<\mathbf{\dot{r}_{CM}}>$ is the body momentum associated with the center of mass $m$
~\cite{Shockley_68,Haus_68}. In particular, it was
shown~\cite{Shockley_68} that the ``hidden linear momentum" has as
quantum mechanical analogue the term $\mathbf{\alpha} \cdot
\mathbf{E}$, where $\mathbf{\alpha}$ are Dirac matrices appearing in
the hamiltonian form $\widehat{H} \psi=i \hbar
\partial \psi/\partial t$, where $\widehat{H}=-ic \hbar \mathbf{\alpha} \cdot
\nabla \cdot$ is the hamiltonian operator (e.g., Ref.~\cite{Sakurai}). Although certainly an important issue, the concept of ``hiddem momentum" needs to be further clarified~\cite{Boyer_05}.

Calkin~\cite{Calkin_71} has shown that the net linear momentum of
any closed stationary system of charges and currents is zero and can
be written:
\begin{equation}\label{eq4}
\mathbf{P}=\int d^3 r \mathbf{r} (\frac{\dot{u}}{c^2})=M
\mathbf{r}_{CM},
\end{equation}
where $u$ is the energy density, $M$ is the total mass, $M=\int d^3
r (u/c^2)$ and $\mathbf{r}_{CM}$ is the radius vector of the center
of mass. He has shown, however, that the linear mechanical momentum
$\mathbf{P}_{ME}$ in a static electromagnetic field is nonzero and
it is given by:
\begin{equation}\label{eq5}
\mathbf{P}_{ME}=-\int d^3 r \rho \mathbf{A}^T.
\end{equation}
Here, $\mathbf{A}^T$ denotes the transverse vector potential given
by $\mathbf{A}^T=(\mu_0/4\pi) \int d^3 r \mathbf{J}/r$. Eq.~\ref{eq5} shows that $\rho \overrightarrow{A}$ is a measure of momentum per unit volume.

Similar conclusion were obtained by Aharonov {\it et al.}~\cite{Aharonov_88}
showing, in particular, that the neutron's electric dipole moment in a
external static electric field $\mathbf{E}_0$ experiences a force
given by $m\mathbf{a}=-(\mathbf{v} \cdot \nabla)(\mathbf{v} \times
\mathbf{E}_0)$. The experimental verification of the Aharonov-Casher effect would confirm total momentum conservation in the
interactions of magnets and charges~\cite{Goldhaber_89}.

Breitenberger~\cite{Breitenberger} discusses thoroughly this
question showing the delicate intricacies behind the subject,
pointing out the conservation of canonical momentum and the
``extremely small" effect of magnetic interactions through the use
of the Darwin's lagrangian derived in 1920~\cite{Darwin_20}.
Boyer~\cite{Boyer_06} applying the Darwin's lagrangian to the system
of a point charge and a magnet have shown that the center-of-energy
has uniform motion. Darwin's lagrangian is correct to the order
$1/c^2$ (remaining Lorentz-invariant) and the procedure to obtain it
eliminates the radiation modes, and thus describing the interaction
of charged particles in the frame on an action-at-a-distance
electrodynamics. It can lead, however, to unphysical
solutions~\cite{Bessonov_99}.

Hnizdo~\cite{Hnizdo_92} have shown that at nonrelativistic
velocities Newton's third law is verified in the interactions between
current-carrying bodies and charged particles because the
electromagnetic field momentum is equal and opposite to the hidden
momentum hold by the current-carrying bodies; the mechanical
momentum of the entire {\it closed} system is conserved. Hnizdo have also
shown that, however, the field angular momentum in a system is not
compensated by hidden momentum and thus the mechanical angular
momentum is not conserved alone, but had to be summed up with the
field angular momentum in order to be a conserved quantity.

In fact, the ``magnetic current force", produced by magnetic charges
that ``flow" when magnetism changes, given by
$\mathbf{f}_m=\varepsilon_0 \mathbf{E} \times (\mathbf{\dot{B}} -
\mu_0 \mathbf{\dot{H}})$\cite{ShockleyJames} is the ``Abraham term"
appearing in the Abraham density force $\mathbf{f}_A$ which differ
from the Minkowsky density force $\mathbf{f}_M$ through the
expression:
\begin{equation}\label{eq6}
\mathbf{f}_A = \frac{\partial }{\partial t} [\mathbf{g}^M - \mathbf{g}^A].
\end{equation}
Here, $\mathbf{^g}^M=[\mathbf{D} \times \mathbf{B}]$ is the Minkowsky's momentum density of the field and $\mathbf{g}^A=[\mathbf{E} \times \mathbf{H}]/c^2$ is the Abraham's momentum density.

\section{Interaction with the vacuum}

Although Newton's third law of motion apparently does not hold in
some situations, it is likely action and reaction always occurs by
pairs and a kind of accounting balance such as
$\mathbf{F}=-\mathbf{F'}$ holds.

According to the Maxwell's theorem, the resultant of $\mathbf{K}$
forces applied to bodies situated within a closed surface $S$ is
given by the integral over the surface $S$ of the Maxwell stresses:
\begin{equation}\label{Eq7}
\int \mathbf{T}(n) dS = \int \mathbf{f} d \Omega = \mathbf{K}.
\end{equation}
Here, $\mathbf{f}$ is the ponderomotive forces density and $d\Omega$
is the volume element. The vector $\mathbf{T}(n)$ under the integral
in the left-hand side (lhs) of the equation is the tension force
acting on a surface element $dS$, with a normal $\mathbf{n}$
directed toward the exterior. In cartesian coordinates, each
component of $\mathbf{T}(n)$ is defined by
\begin{equation}\label{eq8}
T_x(n)=t_{xx} \cos (n,x) + t_{xy} \cos (n,y) + t_{xz} \cos (n,z),
\end{equation}
with similar expressions for $T_y$ and $T_z$. The 4-dimensional
momentum-energy tensor is a generalization of the 3-dimensional
stress tensor $T_{lm}$. If electric charges are inside a conducting
body in vacuum, in presence of electric $E$ and magnetic $H$ fields,
then Eq.~\ref{eq8} must be modified to the form:
\begin{equation}\label{eq9}
\int \mathbf{T}(n) dS - \mathbf{K} = \int \frac{1}{4 \pi c} \left(
\frac{\partial [\mathbf{E} \times \mathbf{H}]}{\partial t} \right) d
\Omega.
\end{equation}
In the right-hand side of the above equation it now appears the
temporal derivative of $\mathbf{G}=\int \mathbf{g}d \Omega$, the
electromagnetic momentum of the field in the entire volume contained
by the surface $S$ (with $\mathbf{g}$ its momentum density).

In the case the surface $S$ is filled with a homogeneous medium
without true charges, Abraham proposed to write instead the
following equation:
\begin{equation}\label{eq10}
\int \mathbf{T}(n) dS = \frac{\partial }{\partial t} \int
\frac{\varepsilon \mu}{4 \pi c} [\mathbf{E} \times \mathbf{H}] d
\Omega,
\end{equation}
with $\varepsilon$ and $\mu$ the dielectric constant of the medium
and its magnetic permeability.

Eq.~\ref{eq10} can be written on the form of a general conservation
law:
\begin{equation}\label{eq11}
\frac{\partial \sigma_{\alpha \beta}}{\partial x_{\beta}} -
\frac{\partial g_{\alpha}}{\partial t} = f_{\alpha}
\end{equation}
where $\alpha=1,2,3$, $\sigma_{\alpha \beta}$ is the stress tensor,
$g_{\alpha}$ is the momentum density of the field, and $f_{\alpha}$
is the total force density. After some algebra this equation can
take the final form:
\begin{equation}\label{eq12}
\frac{\partial \sigma_{\alpha \beta}}{\partial x_{\beta}}=
f_{\alpha}^L + \frac{1}{4 \pi c} \frac{\partial}{\partial t}
[\mathbf{D} \times  \mathbf{B}]_{\alpha} + f'_{m,\alpha}.
\end{equation}
Here, $f'_m$ is the total force acting in the medium (see
Ref.~\cite{Ginzburg76}), $\mathbf{f}^L=\rho_e
\mathbf{E}+\frac{1}{c}[\mathbf{j} \times \mathbf{B}]$ is the Lorentz
force density with $\rho_e$ denoting the charge density and
$\mathbf{j}$ the current density.

Of course, field, matter and physical vacuum form together a closed
system and it is usual to catch the momentum conservation law in the
general form~\cite{Thirring,Landau2,Lee}:
\begin{equation}\label{eq13}
\frac{\partial (T_{\alpha \beta}^{Field} + T_{\alpha \beta}^{Matter}
+ T_{\alpha \beta}^{Vacuum})}{\partial x_{\beta}} = 0.
\end{equation}
Table~\ref{table1} shows the different expressions for the
energy-momentum tensors of Minkowksy, $T_{\alpha,\beta}^M$ and
Abraham, $T_{\alpha,\beta}^A$.

\begin{table*}
\caption{\label{table1} Expressions for the energy-momentum tensors
of Minkowksy $T_{\alpha,\beta}^M$ and Abraham $T_{\alpha,\beta}^A$,
using $i,k=1,2,3,4$; $x_1=x$, $x_2=y$, $x_3=z$, $x_4=ict$. The
Poynting's vector is $\mathbf{S}=[\mathbf{E} \times \mathbf{H}]$ and
the energy for a system at rest is $w=\frac{1}{8 \pi} (\epsilon E^2
+ \mu H^2)$.}
\begin{ruledtabular}
\begin{tabular}{ccc}
Minkowsky        &  Abraham \\ \hline
  $T_{\alpha,\beta}^M= \left(\begin{array}{cc}
                       \sigma_{\alpha,\beta} & -ic\mathbf{g}^M \\
                        -\frac{i}{c}\mathbf{S} & w
                      \end{array} \right)$
  &  $T_{\alpha,\beta}^A=\left( \begin{array}{cc}
                       \sigma_{\alpha,\beta} & -ic\mathbf{g}^A \\
                        -\frac{i}{c}\mathbf{S} & w
                      \end{array} \right) $\\
  $ \mathbf{g}^M = \frac{\epsilon \mu}{c^2}[\mathbf{E} \times \mathbf{H}]$
  & $\mathbf{g}^A = \frac{1}{c^2}[\mathbf{E} \times \mathbf{H}]$ \\
\end{tabular}
\end{ruledtabular}
\end{table*}

The general relation between Minkowski and Abraham momentum, free of
any particular assumption, holding particularly for a moving medium,
is given by
\begin{equation}\label{eq14}
\mathbf{P}^M = \mathbf{P}^A + \int \mathbf{f}^A dt dV.
\end{equation}

For clearness, we shall distinguish the following different parts of
a system: i) the body carrying currents and the currents themselves
(the structure, for short, denoted here by $\mathfrak{K}$), ii) the
fields, and iii) the physical vacuum (or the medium).

On the theoretical ground exposed above, the impulse transmitted to
the structure should be given by the following equation:
\begin{equation}\label{eq15}
\mathbf{P}^{\mathfrak{K}} = \int \mathbf{f}^A dt dV = \mathbf{P}^M -
\mathbf{P}^A.
\end{equation}
Here, $\mathbf{f}^A$ denotes the Abraham's force density~\cite{Abraham1,Abraham2}:
\begin{equation}\label{eq16}
\mathbf{f}^A = \frac{\varepsilon_r \mu_r -1}{4 \pi c} \frac{\partial
[\mathbf{E} \times \mathbf{H}]}{\partial t}.
\end{equation}

\begin{figure}
  \includegraphics[width=3.5 in]{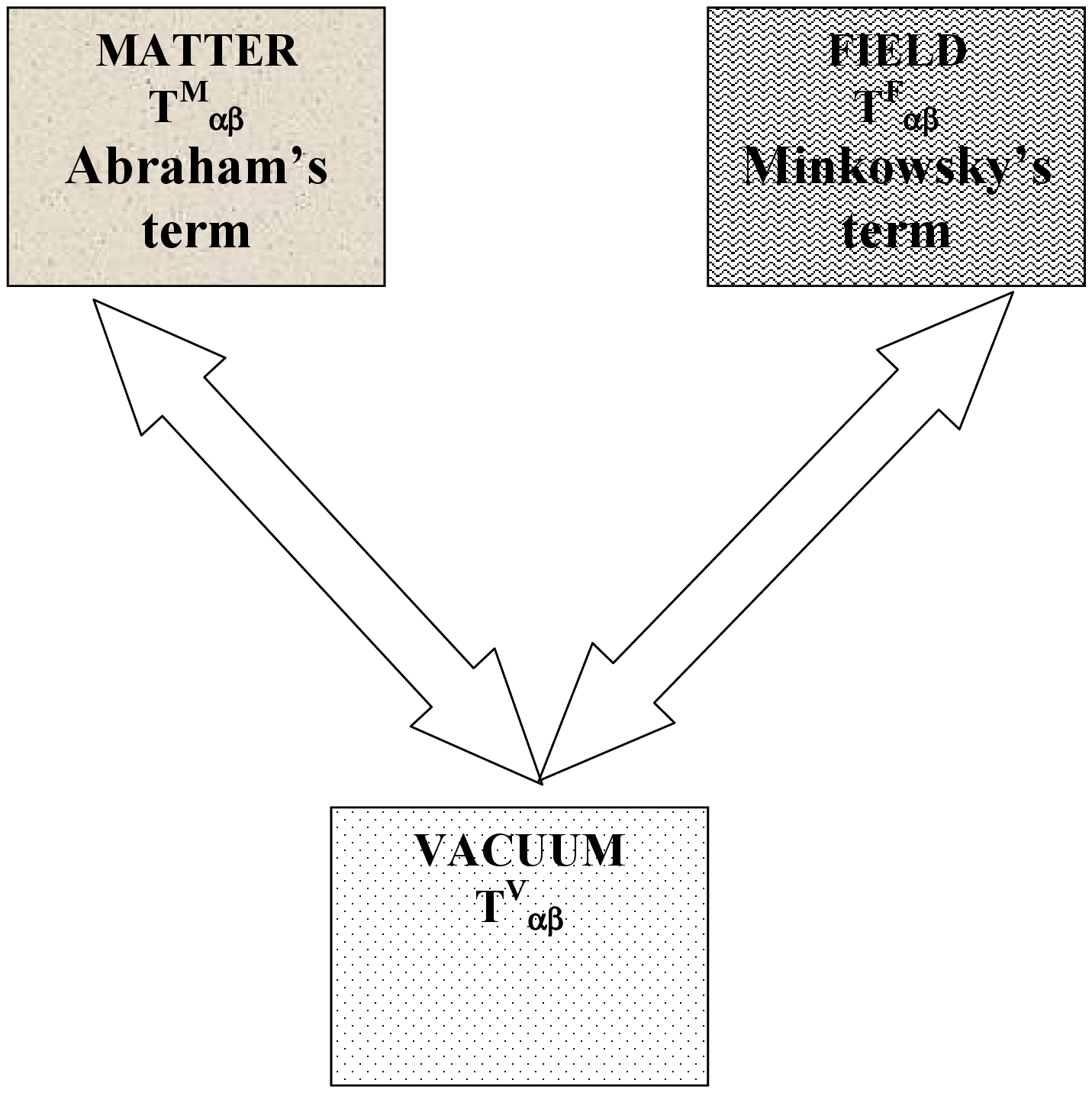}\\
  \caption{Conservation law for the closed system: Matter + Field + Physical Vacuum.}\label{fig1}
\end{figure}

This is in agreement with experimental data ~\cite{Jones} and was
proposed by others ~\cite{Gordon,Tangherlini}. As this force is
acting over the medium, it is expected nonlinearities related to the
behavior of the dielectric to different applied frequencies,
temperature,pressure, and large amplitudes of the electric field
when a pure dielectric response of the matter is no longer
proportional to the electric field (see Ref. ~\cite{Waser} on this
topic).

The momentum conservation law can be rewritten under the general
form (e.g., Ref.~\cite{Ginzburg76}):
\begin{equation}\label{eq17}
\frac{\partial \sigma_{\alpha \beta}}{\partial x_{\beta}} =
f_{\alpha}^L + \frac{1}{4 \pi c} \frac{\partial }{\partial t}
[\mathbf{D} \times \mathbf{B}]_{\alpha} + f_{m,\alpha}^{'},
\end{equation}
with $f_m^´$ denoting the force acting on the medium. The second
term in the r.h.s. of above equation could possible be called {\it
vacuum-interactance} term ~\cite{Clevelance} - in fact, Minkowski
term. Already according to an interpretation of Einstein and Laub
~\cite{Laub}, the integration of above equation over all space, the
derivative over stress tensor gives a null integral and the Lorentz
forces summed over all the universe must be balanced by the quantity
$\int_{\infty} \varepsilon_0 \mu_0 \frac{\partial [\mathbf{E} \times
\mathbf{H}]}{\partial t} dV$ in order Newton's third law be
preserved. It is important to remark that the field momentum $[\mathbf{D} \times \mathbf{B}]$ is equivalent to $\rho \mathbf{A}$, the first term is related to the stress-tensor representation, while the second one is related to the ``fluid-flow" representation~\cite{Carpenter_89}.

As is well known, Maxwell's classical theory introduces the idea of
a real vacuum medium. After being considered useless by Einstein's
special theory of relativity, the ``ether" (actually replaced by the
term {\it vacuum} or {\it physical vacuum}) was rehabilitated by
Einstein in 1920 ~\cite{Einstein22}. In fact, general theory of
relativity describes space with physical properties by means of ten
functions $g_{\mu \nu}$ (see also ~\cite{Ginzburg87}). According to
Einstein, \begin{quote} The ``ether" of general relativity is a
medium that by itself is devoid of {\it all} mechanical and
kinematic properties but at the same time determines mechanical (and
electromagnetic) processes.
\end{quote}

Dirac felt the need to introduce the idea of ``ether" in quantum
mechanics ~\cite{Dirac}. In fact, according to quantum field theory,
particles can condense in vacuum giving rise to space-time dependent
macroscopic objects, for example, of ferromagnetic type. Besides,
stochastic electrodynamics have shown that the vacuum contains
measurable energy called zero-point energy (ZPE) described as a
turbulent sea of randomly fluctuating electromagnetic field. Quite
interestingly, it was recently shown that the interaction of atoms
with the zero-point field (ZPF) guarantees the stability of matter
and, in particular, the energy radiated by an accelerated electron
in circular motion is balanced by the energy absorbed from the ZPF
~\cite{kozlowski}. An attempt to replace a field by a finite number
of degrees of freedom was done by Pearle~\cite{Pearle_71}. In this
theory $N$ particles were supposed do not interact directly with
each other, but interact directly with a number of dynamical
variables (called the ``medium") carrying the "information" from one
particle to another.

Graham and Lahoz made three important
experiments~\cite{Lahoz1,Lahoz2,Lahoz3}. While the first experiment
provided an experimental observation of Abraham force in a
dielectric, the second one provided a measurement of a reaction
force which appear in magnetite. The third one provided the first
evidence of free electromagnetic angular momentum created by
quasistatic and independent electromagnetic fields $E$ and $B$ in
the vacuum ~\cite{Note1}. Whereas the referred paper by Lahoz
provided experimental evidence for Abraham force at low frequency
fields, it still remains to gather evidence of its validity at
higher frequency domain, although some methods are presently
outlined~\cite{Antoci}.

All this is known since a long time and we only try to put more
clear the theoretical framework, that only needs to be
experimentally tested for proof of principles.

In view of the above, we will write the ponderomotive force density
acting on the composite body of arbitrarily large mass (formed by
the current configuration and its supporting structure) in the form
\begin{equation}\label{eq18}
\rho \frac{d \mathbf{V}}{dt} = \rho_c \mathbf{E} + [\mathbf{J}
\times \mathbf{B}] + \nabla \cdot \mathbf{T} +
\frac{\partial}{\partial t} \left( \varepsilon_0 \mu_0 [\mathbf{E}
\times \mathbf{H}] \right).
\end{equation}

Hence, the composite body is acted on by Minkowski force in such a
way that
\begin{equation}\label{eq19}
M\mathbf{V} = -\mathbf{G}^M + \mathbf{G}^A.
\end{equation}
The Minkowski momentum is transferred only to the field in the
structure and not to the structure and the field in the
medium~\cite{Skobeltsyn,Ginzburg76,Lahoz3}. In summary, to move a
spacecraft forward, the spacecraft must push ``something" backwards;
and this ``something" might be the physical vacuum. This effect was shown to be made feasible, the Abraham's force representing the reaction of the physical vacuum fluctuations to the motion of dielectric fluids in crossed electric and magnetic fluids communicating to matter velocities of the order of 50 nm$/$s~\cite{Feigel_04}, although this result was contested by van Tiggelen {\it et al.}~\cite{Tiggelen_04}.

The exploration of these ideas to propel a spacecraft has been advanced in the literature, e.g., see Refs.~\cite{Taylor,Trammel,Brito,Forward2004}.

\section{Deducing the linear momentum for a material body on the basis of statistical physics}

When two bodies of matter collide, the repulsive force on them is
equal whenever no dissipative process is at stake. When a ball
rebound on the floor it has the same total mechanical energy before
and after the collision, except for a loss term which is due to the
fact that the bodies have internal structure. At a microscopical
level, bodies are aggregates of molecules. When the body collides,
molecules gain an internal (random) kinetic energy. Macroscopically
this generates heat, and therefore raises the system entropy. In
global terms, some fraction of heat does not return to the
particle's collection constituting the ball and the entropy of the
universe ultimately increases.

Let's consider an isolated material body composed by a great number
of macroscopic particles (let's say $N$) possessing an internal
structure with a great number of degrees of freedom (to validate the
entropy concept) with momentum $\mathbf{p}_i$, energy $E_i$ and with
intrinsic angular momentum $\mathbf{J}_i$, all constituted of
classical charged particles with charge $q_i$ and inertial mass
$m_i$. Using the procedure outlined in
Refs.~\cite{Pinheiro:02,Pinheiro:04} we can show that the entropy
gradient in momentum space is given by
\begin{equation}\label{eq20}
\mathbf{p}_i = m_i \mathbf{v}_e + q_i \mathbf{A} + m_i
[\mathbf{\omega} \times \mathbf{r}_i] -m_i T_i \frac{\partial
\overline{S}}{\partial \mathbf{p}_i}.
\end{equation}
It was assumed that all particles have the same drift velocity and
they turn all at the same angular velocity $\omega$. The center of
mass of the body moves with the same macroscopic velocity and the
body turns at the same angular velocity~\cite{Landau2}. The last term
of Eq.~\ref{eq20} represents the gradient of the entropy in a
nonequilibrium situation and $\overline{S}$ is the transformed
function defined by:
\begin{widetext}
\begin{equation}\label{eq21}
\overline{S} = \sum_{i=1}^{N} \left\{ S_i \left[ E_i -
\frac{p_i^2}{2m_i} - \frac{J_i^2}{2I_i} -q_i V_i + q_i (\mathbf{A}_i
\cdot \mathbf{v_i})] + (\mathbf{a \cdot \mathbf{p}_i}) + \mathbf{b}
\cdot ([\mathbf{r}_i \times \mathbf{p}_i] + \mathbf{J}_i) \right]
\right\},
\end{equation}
\end{widetext}
where $\mathbf{a}$ and $\mathbf{b}$ are Lagrange multipliers.

Whenever the system is in thermodynamic equilibrium the canonical
momentum is obtained for each composing particle:
\begin{equation}\label{eq22}
\mathbf{p}_i = \mathbf{p}_{rel} + m_i [\mathbf{\omega} \times
\mathbf{r}_i] + q_i \mathbf{A}_i.
\end{equation}
Otherwise, when the system is subjected to forced constraints in
such a way that entropic gradients in momentum space do exist, then
a new expression for the particle momentum must be taken into
account, that is, Eq.~\ref{eq20}.

Summing up over all the constituents particles of a given
thermodynamical system pertaining to the same aggregate (e.g., body
or Brownian particle), we obtain:
\begin{equation}\label{eq24}
\mathbf{P} = M \mathbf{v}_e + \sum_i m_i [\mathbf{\omega} \times
\mathbf{r}_i] + Q \mathbf{A} - \sum_i m_i T_i \frac{\partial
\overline{S}}{\partial \mathbf{p}_i}.
\end{equation}
To simplify we can assume that all the particles inside the system
have the same random kinetic energy, $T_i=\zeta$:
\begin{equation}\label{eq25}
\mathbf{P} = M \mathbf{v}_e + \sum_i m_i [\mathbf{\omega} \times
\mathbf{r}_i] + Q \mathbf{A} - \zeta \sum_i \frac{\partial
\overline{S}_{ne}}{\partial \dot{r}_i},
\end{equation}
where by $\overline{S}_{ne}$ we denote the entropy when the system
is in a state out of equilibrium. The first term on the right-hand
side is the bodily momentum associated with motion of the center of
mass $M$; the second term represents the rotational momentum; the
third is the momentum of the joint electromagnetic field of the
moving charges~\cite{Fowles_80,Scanio_75}; finally, the last term is
a new momentum term, that can be physically understood as a kind of
``entropic momentum" since it is ultimately associated to the
information exchanged with the medium on the the physical system
viewpoint (e.g., momentum that eventually is radiated by the charged
particle). Lorentz's equations don't change when time is reversed,
but when retarded potentials are applied the time delay of
electromagnetic signals on different parts of the system do not
allow perfect compensation of internal forces, introducing
irreversibility into the system~\cite{Ritz_08}. This is always true
whenever there is time-dependent electric or/and magnetic
fields~\cite{Jefimenko_1}. Cornish~\cite{Cornish_86} obtained a
solution of the equation of motion of a simple dumbbell system held
at fixed distance and have shown that the effect of radiation
reaction on an accelerating system induces a self-accelerated
transverse motion. Obara and Baba~\cite{Baba_00} have discussed the electromagnetic propulsion of a electric dipole system and they have shown that the propulsion effect results from the delay action of the static and inductive near-field created by one electric dipole on the other.
These are examples of irreversible (out of
equilibrium) phenomena that do not comply with action-reaction law.

At this stage, we can argue that the momentum is always a conserved
quantity provided that we add the right term, making Newton's third
law verified. This apparent ``missing symmetry" might result because
matter alone does not form a closed system, and we need to include
the physical vacuum in order symmetry be restored. So, when we have
two systems $1$ and $2$ interacting via some kind of force field
$\mathbf{F}$, the reaction from the vacuum must be included as a
sort of bookkeeping device:
\begin{equation}\label{eq26}
\mathbf{F}_{12}^{matter}=-\mathbf{F}_{21}^{matter} +
\mathbf{F}^{vacuum}.
\end{equation}
We may assume the existence of a physical vacuum probably well
described by a spin-0 field $\phi(x)$ whose vacuum expectation value
is not zero:
\begin{equation}\label{eq27}
\text{vacuum} \sim \phi(x),
\end{equation}
and at its lowest-energy state to have zero 4-momentum, $k_{\mu}=0$
(e.g., Ref.~\cite{Lee}).

This new state out of equilibrium can be constrained by applying an
external force on the system (e.g., set all system into rotation
about its central axis at the same angular velocity
$\mathbf{\omega}$).

It was shown that the entropy must increase with a small
displacement from a previous referred state ~\cite{Lavenda,Landau2}.
Considering that the entropy is proportional to the logarithm of the
statistical weight $\Omega \propto exp(\overline{S}/k_B)$ and
considering that $\overline{S}=\overline{S}_{eq}+\overline{S}_{ne}$
we can expect an increase of the nonequilibrium entropy
$\overline{S}_{ne}$ with a small increase of the i$th$ particle's
velocity $\mathbf{v}_i=\mathbf{\dot{r}_i}$, since with an increase
of particle's speed (although in random motion) the entropy must increases altogether.
Therefore, we must always have:
\begin{equation}\label{eq28}
T \frac{\partial \overline{S}_{ne}}{\partial \mathbf{\dot{r}_i}} \geq
0, \forall i=1,...N.
\end{equation}
In conditions of mechanical equilibrium the equality must hold, otherwise condition ~\ref{eq28} can be considered a {\it
universal criterium of evolution}. Considering that the entropy is
an invariant~\cite{Rengui} there is no extra similar term when the
momentum is transferred to another inertial frame of reference.

Quite withstanding, there is an important theorem derived by Baierlin~\cite{Baierlein}
showing that the Gibbs entropy for a system of free particles with
kinetic energy $K$, density $\rho$ and absolute temperature $T$,
$S(K,\rho,T)$, is greater than the entropy associated to the same
system subject to arbitrary velocity-independent interactions $V$,
$S(K+V,\rho,T)$, such as $S(K+V,\rho,T) \leq S(K,\rho,T)$.

At the electromagnetic level, Maxwell conceived a dynamical model of
a vacuum with hidden matter in motion. As it is well-known,
Einstein's theory of relativity eradicated the notion of ``ether"
but later revived its interest in order to give some physical mean
to $g_{ij}$. Minkowski obtained as a mathematical consequence of the
Maxwell's mechanical medium that the Lorentz's force should be
exactly balanced by the divergence of the Maxwell's tensor in vacuum
$T_{vac}$ minus the rate of change of the Poynting's vector:
\begin{equation}\label{eq29}
\rho \mathbf{E} + \mu_0 [\mathbf{J} \times \mathbf{H}] = \nabla
\cdot \mathrm{T}_{vac} -\frac{\partial }{\partial t} \varepsilon_0
\mu_0 [\mathbf{E} \times \mathbf{H}].
\end{equation}
Einstein and Laub remarked~\cite{Laub} that when Eq.~\ref{eq8} is
integrated all over the entire Universe the term $\nabla \cdot
\mathrm{T}_{vac}$ vanish which means that the sum of all Lorentz
forces in the Universe must be equal to the quantity $\int_{\infty}
\varepsilon_0 \mu_0 \partial / \partial t [\mathbf{E} \times
\mathbf{H}] dv$ in order to comply with Newton's third law (see
Ref.~\cite{Lahoz}). But, this long range force depends on the
constant of gravitation $G$. Einstein accepted the Faraday' point of
view of the reality of fields, and this gravitational field
according to him would propagate all over the entire space without
loss, locally obeying to the action-reaction law. But nothing can
reassure us that the propagating wave through the vacuum will be
lost at infinite distances~\cite{Brillouin_70}.
Poincar\'{e}~\cite{Poincare_01} also argues about the possible
dissipation of the action on matter due to the absorption of the
propagating wave in the context of Lorentz's theory.

By Noether's theorem, energy conservation is related to
translational invariance in time ($t \to t +a$) and momentum
conservation is related to translational invariance in space ($r_i
\to r_i + b_i$). This important theorem thus implies that the law of
conservation of momentum (not equivalent to the
action-equals-reaction principle) is always valid, while the law of
action and reaction does not always holds, as shown in the previous
examples.

Some kind of relationship must therefore exists between entropy and
Newton's third law, as it was through the combined equation with the
first and second law of thermodynamics that our main result were
obtained. This idea was verified recently through a standard
Smoluchowski approach and on Brownian dynamic computer simulation of
two fixed big colloidal particles in a bath of small Brownian
particles drifting with uniform velocity along a given direction. It
was shown that, in striking contrast to the equilibrium case, the
nonequilibrium effective force violates Newton's third law, implying
the presence of nonconservative force showing a strong
anisotropy~\cite{Likos_03}. This result reminds our Eq.~\ref{eq26}.

\section{Is it verified the action-equals-reaction law in a thermodynamical system out-of-equilibrium ?}

\begin{figure}
  \includegraphics[width=3.0 in, height=4.0 in]{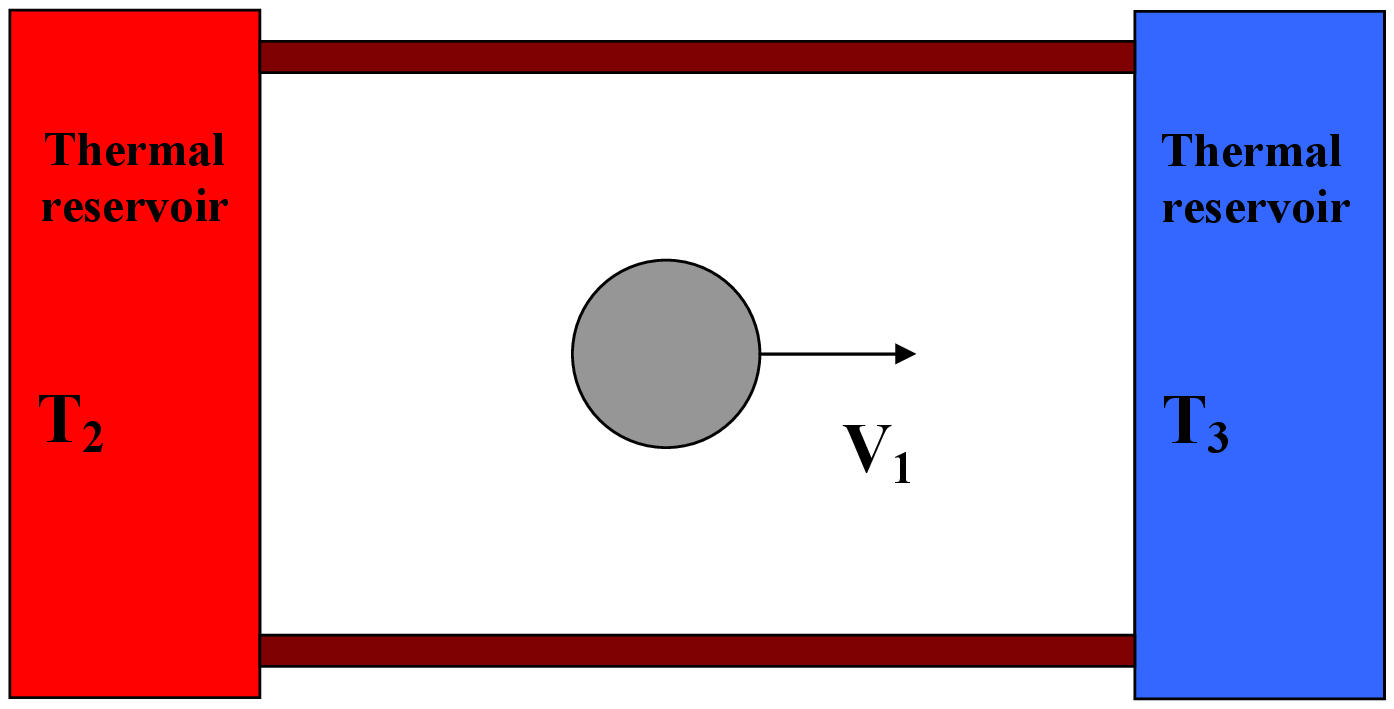}\\
  \caption{Schematic of the self-accelerated device.}\label{fig2}
\end{figure}

The maximizing entropy procedure proposed in Ref.~\cite{Pinheiro:02,Pinheiro:04} suggest the following ``gedankenexperiment". This problem bears some resemblance to the Leo Szilard's thermodynamic engine with one-molecule fluid (e.g., Ref.~\cite{Leff}), although we are not concerned here with neguentropy issues.

Let us suppose a system consisting of a spherical body made of
$N$ number of particles closed in a box and moving along one direction (see Fig.~\ref{fig2}). The
left side is at temperature $T_2$, the right side is at
temperature $T_3$, while the body particle itself is at temperature $T_1$
(and in equilibrium with their photonic environment).
Furthermore, we assume that both surfaces and the body particle
are all thermal reservoirs, and hence their respective temperatures do not change.
Let us suppose that the onset of nonequilibrium dynamics can be forced by some means in the previously described device.
When the particle collides with the top its momentum varies according to:
\begin{equation}\label{eq30}
\delta p_{\upharpoonleft}=-mv_1" +mv_1 + (T_3-T_1)
\partial_v S.
\end{equation}
Here, $\partial_v S$ denotes the (nonequilibrium) entropy gradient in velocity space.
After the collision the particle goes back to hit the right surface
at temperature $T_3$. The momentum variation after the second
collision is given by:
\begin{equation}\label{eq31}
\delta p_{\downharpoonleft}= mv_1´ - mv_1" + (T_2 - T_1)
\partial_v S.
\end{equation}
We assume that the body attain thermal equilibrium with the
environment (which must remain at constant temperature $T_1$) fast enough before the next hit against the wall of the thermal reservoir. The total balance after a complete loop, back and forth, is given by
\begin{equation}\label{eq32}
\delta p_{\downharpoonleft} = - \delta p_{\upharpoonleft}
-\partial_v S(T_2 + T_3 - 2 T_1)=-\delta p_{\upharpoonleft}-\Delta \zeta \nabla_v S.
\end{equation}
To make it more clear, we might write Eq.~\ref{eq32} under the form
\begin{equation}\label{eq33}
\delta p_{\downharpoonleft}=-\delta p_{\upharpoonright}-\delta p_{\upharpoonright}^{is},
\end{equation}
where we denote by $\delta p_{\upharpoonright}^{is}\equiv \Delta \zeta \nabla_v S$, the change in momentum by the physical vacuum (others, would call ``inertial space").
Therefore, it is clear from the above analysis that
{\it in systems out of equilibrium Newton's third law is not
verified}. The conservation of canonical momentum, however, is well
verified, as it must be according to Noether's theorem. Otherwise,
when the temperatures are equal for all thermal bath in contact,
such as $T_1=T_2=T_3$, Newton's third law is verified:
\begin{equation}\label{eq33}
\delta p_{\downharpoonleft} = - \delta p_{\upharpoonleft}.
\end{equation}
In the frame of nonlinear dynamics and statistical approach Denisov has shown~\cite{Denisov_02} that a rigid shell and a nucleus with internal dynamic asymmetric can perform self unidirectional propulsion. It seems now certain that depletion forces between two fixed big colloidal particles in a bath of small particle exhibits nonconservative and strongly anisotropic forces that violate action-reaction law~\cite{Likos_03} (see also Ref.~\cite{Wang_02}). Also, internal Casimir forces between a circle and a plate in nonequilibrium situation violates Newton's law~\cite{Soto_08}.

\section{Conclusion}

The purpose of this study is to examine how the action-reaction law is faced in the literature in the domain of mechanics, electrodynamics and statistical mechanics, and to offer a methodological approach in order to tackle the fundamental aspect of the problem suggesting that a third part should be included in the analysis of forces, what we called here for the sake of conciseness, the physical vacuum.
Furthermore, a general procedure lead us
to an expression for the general linear (canonical) momentum of a
body-particle in the framework of statistical mechanics. Theoretical
arguments and numerical computations suggest that Newton's third law
is not verified in out-of-equilibrium systems due to an additional
entropic gradient term present in the particle's canonical momentum.
Although Noether's theorem guaranty the conservation of canonical
momentum, the actions-equal-reaction principle can be restored in
nonequilibrium conditions only if a new force term representing the
action of the medium on the particles is taken into account.

\bibliographystyle{amsplain}
\bibliography{Doc2}

\begin{thebibliography}{99}
\bibitem{Note} ``It is not enough an indifferent middle
agent, like the center of opposed forces acting on the lever; it is
necessary an agent that, by its own virtues, measure and control the
force" (translated by the author)
\end{thebibliography}

\end{document}